\documentclass[11pt]{article}
\usepackage{amsmath}
\usepackage{amsfonts}
\usepackage{amssymb}
\usepackage[usenames]{color}
\usepackage{theorem}

\def\l{\lambda}
\def\p{\partial}
\def\e{\mathrm{e}}

\newcommand{\be}{\begin{equation}}
\newcommand{\ee}{\end{equation}}
\newcommand{\bea}{\begin{eqnarray}}
\newcommand{\eea}{\end{eqnarray}}
\newcommand{\beaa}{\begin{eqnarray*}}
\newcommand{\eeaa}{\end{eqnarray*}}

\newcommand{\nn}{\nonumber}
\renewcommand{\d}{\mathrm{d}}
\begin{document}
\title{Non-Hamiltonian generalizations of the dispersionless 2DTL hierarchy}
\author{
L.V. Bogdanov\thanks
{L.D. Landau ITP RAS,
Moscow, Russia, e-mail
leonid@landau.ac.ru}}
\maketitle
\begin{abstract} 
We consider two-component integrable generalizations of  the 
dispersionless 2DTL hierarchy connected with 
non-Hamiltonian vector fields, similar to the Manakov-Santini hierarchy generalizing 
the dKP hierarchy.
They form a one-parametric family 
connected by hodograph type transformations. Generating
equations and Lax-Sato equations are introduced, a dressing scheme based on the
vector nonlinear Riemann problem is formulated. 
The simplest two-component generalization of the dispersionless 2DTL equation is derived,
its differential reduction analogous to the Dunajski interpolating system is presented.
A symmetric two-component generalization of the dispersionless elliptic 
2DTL equation is also constructed.
\end{abstract}
\section{Introduction}
Recently S.V. Manakov and P.M. Santini introduced a two-component system   
generalizing
the dispersionless KP equation to the case of 
non-Hamiltonian vector fields in the Lax pair \cite{MS06,MS07}, 
\bea
u_{xt} &=& u_{yy}+(uu_x)_x+v_xu_{xy}-u_{xx}v_y,
\nn\\
v_{xt} &=& v_{yy}+uv_{xx}+v_xv_{xy}-v_{xx}v_y,
\label{MSeq}
\eea
and the Lax pair is
\bea
&&
\partial_y\mathbf{\Psi}=((\l-v_{x})\partial_x - u_{x}\partial_\l)\mathbf{\Psi},
\nn\\
&&
\partial_t\mathbf{\Psi}=((\l^2-v_{x}\l+u -v_{y})\partial_x
-(u_{x}\l+u_{y})\partial_\l)\mathbf{\Psi},
\label{MSLax}
\eea
where $u$, $v$ are functions of $x$, $y$, $t$, and $\l$ 
plays a role of a spectral variable.
For $v=0$  the system (\ref{MSeq}) reduces to the dKP
(Khohlov-Zabolotskaya) equation
\be
u_{xt} = u_{yy}+(uu_x)_x.
\label{dKP-eq}
\ee
Respectively, the reduction $u=0$ gives an equation 
\cite{Pavlov03} 
\begin{equation}
 v_{xt} = v_{yy}+v_xv_{xy} - v_{xx}v_y.
\label{Pavlov}
\end{equation}
The  hierarchy related to this system was studied in \cite{LVB1,LVB2}. 
It was demonstrated
that the Manakov-Santini hierarchy represents a case N=1 of a general (N+1)-component
hierarchy. This general hierarchy is connected with commutativity of (N+1)-dimensional 
vector vields containing a derivative with respect to the spectral variable, with the 
coefficients of vector fields meromorphic in the complex plane 
of the spectral variable and having a pole only at one point
(e.g., infinity, compare the Lax pair (\ref{MSLax})). In this sense the Manakov-Santini 
hierarchy is a
two-component one-point hierarchy, and generalizations of the 
dispersionless 2DTL hierarchy we are going to consider in this paper
represent a two-component two-point case, when vector 
fields have poles at two points (say, zero and infinity). 

Our starting point is the
formalism of the works \cite{BDM07,LVB1,LVB2}, which we transfer to the two-point case
having in mind the representation of the dispersionless 2DTL hierarchy given in 
\cite{TT91,TT95} 
{and the results of the recent work \cite{MS09},
in which the dressing, the Cauchy problem 
and the behavior of solutions of the dispersionless 2D Toda equation were studied.
}
We introduce generating equations and Lax-Sato equations and develop 
a dressing scheme based on vector
nonlinear Riemann problem. We discover one-parametric freedom in 
generalizing the dispersionless 2DTL hierarchy, 
and describe hodograph type transformations connecting
different generalizations. 

The simplest two-component generalization of the dispersionless 2DTL equation reads
\bea
&&
(\mathrm{e}^{-\phi})_{tt}=m_t\phi_{xy}-m_x\phi_{ty},
\nn\\
&&
m_{tt}\mathrm{e}^{-\phi}=m_{ty}m_x-m_{xy}m_t,
\label{gen2DTL}
\eea
and the Lax pair is
\bea
&&
\partial_x\mathbf{\Psi}=\left((\l+ \frac{m_x}{m_t})\partial_t - 
\l(\phi_t \frac{m_x}{m_t}-\phi_x)\partial_\l\right)\mathbf{\Psi},
\nn\\
&&
\partial_y\mathbf{\Psi}=\left(\frac{1}{\l}\frac{\mathrm{e}^{-\phi}}{m_t}\partial_t +
\frac{(\mathrm{e}^{-\phi})_t}{m_t}\partial_\l\right)\mathbf{\Psi}
\label{Laxgen2DTL}
\eea
(the derivation is given below).
For $m=t$ the system (\ref{gen2DTL}) reduces to the dispersionless 2DTL equation
\bea
(\mathrm{e}^{-\phi})_{tt}=\phi_{xy},
\label{d2DTL}
\eea
Respectively, the reduction $\phi=0$ gives an equation 
\cite{Pavlov03}
\beaa
m_{tt}=m_{ty}m_x-m_{xy}m_t.
\eeaa 
System (\ref{gen2DTL}) doesn't preserve the symmetry of the dispersionless 2DTL equation 
with respect 
to $x$, $y$ variables, however, we also introduce a symmetric generalization of
the d2DTL equation.
\section{Generalized dispersionless 2DTL hierarchy} 
We generalize a picture
of the dispersionless 2DTL hierarchy given by Takasaki and Takebe \cite{TT91,TT95},
{taking into account the results of the recent work \cite{MS09}},
to the case of non-Hamiltonian vector fields, similar to the Manakov-Santini hierarchy,
which generalizes the dispersionless KP hierarchy \cite{MS06,MS07,LVB1,LVB2}.
We consider formal series
\bea
&&
\Lambda^\text{out}=\ln\lambda+\sum_{k=1}^{\infty}l^+_k\lambda^{-k},\quad
\Lambda^\text{in}=\ln\lambda+\phi+\sum_{k=1}^{\infty}l^-_k\lambda^{k},
\label{Lform}\\
&&
M^\text{out}=M_0^\text{out} + \sum_{k=1}^{\infty}m^+_k\e^{-k\Lambda^{+}},\quad
M^\text{in}=M_0^\text{in} + m_0 +\sum_{k=1}^{\infty}m^-_k\e^{k\Lambda^{-}},
\label{Mform}\\
&&
M_0=t+x\e^\Lambda+y\e^{-\Lambda}+\sum_{k=1}^{\infty}x_k\e^{(k+1)\Lambda}+
\sum_{k=1}^{\infty}y_k\e^{-(k+1)\Lambda},\nn
\eea
where  $\lambda$ is a spectral variable. Usually we
suggest that `out' and `in' components of the series define the functions
outside and inside the unit circle in the complex plane of the variable
$\lambda$ respectively, 
with $\Lambda^\text{in}-\ln\lambda$, $M^\text{in}-M_0^\text{in}$ analytic
in the unit disc, and 
$\Lambda^\text{out}-\ln\lambda$, $M^\text{out}-M_0^\text{out}$ analytic
outside the unit disc and decreasing at infinity.
For a function on the complex plane, having a discontinuity on the unit circle, 
by `in' and `out' components we mean the function inside and outside the unit disc.
For two-component series we observe a natural
convention $(AB)^\text{in}=A^\text{in}B^\text{in}$, 
$(AB)^\text{out}=A^\text{out}B^\text{out}$,
which corresponds to multiplication of respective functions on the complex plane.
The coefficients of the series $\phi$, $m_0$, $l^{\pm}_k$, $m^{\pm}_k$
are functions of times $t$, $x_n$, $y_n$. 
Usually for simplicity we suggest that only finite number of $x_k$, $y_k$
are not equal to zero.

Generalized dispersionless 2DTL hierarchy is defined by the generating relation
\be
((J_0)^{-1}\d\Lambda\wedge \d M)^\text{out}=((J_0)^{-1}\d\Lambda\wedge \d M)^\text{in},
\label{genToda}
\ee
which may be considered as a continuity condition on the unit circle for the differential
two-form (or just in terms of formal series),
where $J_0$ is a determinant of Jacobi type matrix $J$,
\beaa
J=
\begin{pmatrix}
\l\p_\l \Lambda & \p_t \Lambda\\
\l\p_\l M       & \p_t M
\end{pmatrix},
\eeaa
$J_0^\text{out}=1+O(\lambda^{-1})$, $J_0^\text{in}=1+\p_t m_0 + O(\lambda)$,
and we suggest that $J_0\neq 0$;
the differential $\d$ is given by 
\be
\d f=\p_\lambda f \d\lambda + \p_t f\d t+
\sum_{k=1}^{\infty}\frac{\p f}{\p x_k}\d x_k+
\sum_{k=1}^{\infty}\frac{\p f}{\p y_k}\d y_k.
\label{diff}
\ee
As a result of a continuity condition, 
the coefficients of the differential two-form in the generating relation
(\ref{genToda}) are {\em meromorphic}.

First we will give a direct derivation of the Lax-Sato equations of 
generalized two-component d2DTL 
hierarchy from the generating relation (\ref{genToda}). 
It is also possible to give a derivation based on an 
intermediate general statement about
linear operators of the hierarchy, similar to the works \cite{BDM07}, \cite{LVB1},
but here we prefer to demonstrate a more straightforward way of exploiting
the generating relation (\ref{genToda}). 

Taking a term of the generating
relation containing $\d \l\wedge\d x_n$, we get
\beaa
&&
((J_0)^{-1}(\p_\l\Lambda  \p_n^+ M-
\p_\l M \p_n^+ \Lambda)\d \l\wedge\d x_n)^\text{out}
\\&&
\quad
=((J_0)^{-1}(\p_\l\Lambda  \p_n^+ M-
\p_\l M \p_n^+ \Lambda)\d \l\wedge\d x_n)^\text{in},
\eeaa
where we introduce a notation 
$\p^+_n=\frac{\p}{\p x_n}$, $\p^-_n=\frac{\p}{\p y_n}$. Thus, taking into account
(\ref{Lform}), (\ref{Mform}), we come to the conclusion that the functions
\bea
A^+_n=\lambda(J_0)^{-1}(\p_\l\Lambda  \p_n^+ M-
\p_\l M \p_n^+ \Lambda)
\label{linA}
\eea
are polynomials, and they can be expressed by the formula
\bea
A^+_n=((J_0)^{-1}(\l\p_\l{\Lambda}) \e^{(n+1)\Lambda})^\text{out}_+,
\label{defA}
\eea
where the subscripts ${}_+$, ${}_-$ denote projection operators, 
$(\sum_{-\infty}^{\infty}u_n p^n)_+=\sum_{n=0}^{\infty}u_n p^n$,
$(\sum_{-\infty}^{\infty}u_n p^n)_-=\sum_{-\infty}^{n=-1}u_n p^n$.
In a similar way,  taking a term of the generating
relation containing $\d t\wedge\d x_n$, we conclude that the functions
\bea
B^+_n=(J_0)^{-1}(\p_t\Lambda  \p_n^+ M-
\p_t M \p_n^+ \Lambda)
\label{linB}
\eea
are also polynomials, and they can be expressed by the formula
\bea
B^+_n=((J_0)^{-1}(\p_t{\Lambda}) \e^{(n+1)\Lambda})^\text{out}_+,
\label{defB}
\eea
Resolving (\ref{linA}), (\ref{linB}) as linear equations with respect to
$\p_n^+ \Lambda$, $\p_n^+ M$, we obtain Lax-Sato equations for the times
$x_n$,
\beaa
\partial^+_n
\begin{pmatrix}
\Lambda\\
M
\end{pmatrix}=
(A^+_n\p_t-B^+_n \l\p_\l)
\begin{pmatrix}
\Lambda\\
M
\end{pmatrix}
\eeaa
Taking the terms of the generating
relation containing $\d \l\wedge\d y_n$, $\d t\wedge\d y_n$, 
we obtain Lax-Sato equations for the times
$y_n$,
\beaa
\partial^-_n
\begin{pmatrix}
\Lambda\\
M
\end{pmatrix}=
(A^-_n\p_t-B^-_n \l\p_\l)
\begin{pmatrix}
\Lambda\\
M
\end{pmatrix},
\eeaa
where
\beaa
&&
A^-_n=((J_0)^{-1}(\l\p_\l{\Lambda}) \e^{-(n+1)\Lambda})^\text{in}_-,
\\
&&
B^-_n=((J_0)^{-1} (\p_t{\Lambda}) \e^{-(n+1)\Lambda})^\text{in}_-.
\eeaa
The compatibility of the flows defined by the Lax-Sato equations can be proved
similar to the case of Dunajski hierarchy \cite{BDM07}, see also \cite{LVB2}.
In explicit form, a complete set of Lax-Sato equations reads
\bea
&&
\left(
\frac{\partial^+_n}{n+1}
-\left(\frac{\l(\e^{(n+1)\Lambda})_\l}
{\{\Lambda,M\}}
\right)^\text{out}_+\p_t
+ 
\left(\frac{(\e^{(n+1)\Lambda})_t}
{\{\Lambda,M\}}
\right)^\text{out}_+
\l\p_\l
\right)
\begin{pmatrix}
\Lambda\\
M
\end{pmatrix}=0,\qquad
\label{Hi1}
\\
&&
\left(
\frac{\partial^-_n}{n+1}
+\left(\frac{\l(\e^{-(n+1)\Lambda})_\l}
{\{\Lambda,M\}}
\right)^\text{in}_-\p_t
- 
\left(\frac{(\e^{-(n+1)\Lambda^-})_t}
{\{\Lambda,M\}}
\right)^\text{in}_-
\l\p_\l
\right)
\begin{pmatrix}
\Lambda\\
M
\end{pmatrix}=0,\qquad
\label{Hi2}
\eea
where the definition of the Poisson bracket is $\{f,g\}=\l(f_\l g_t-f_t g_\l)$.
Lax-Sato equations for the times $x=x_1$, $y=y_1$,
$\partial^+_1=\p_x$, $\partial^-_1=\p_y$,
\beaa
&&
\partial_x\mathbf{\Psi}=\left((\l+ (m_1^+)_t-l_1^+)\partial_t - 
\l l_1^+\partial_\l\right)\mathbf{\Psi},
\\
&&
\partial_y\mathbf{\Psi}=\left(\frac{1}{\l}\frac{\mathrm{e}^{-\phi}}{m_t}\partial_t +
\frac{(\mathrm{e}^{-\phi})_t}{m_t}\partial_\l\right)\mathbf{\Psi},
\eeaa
where $\mathbf{\Psi}=\begin{pmatrix}
\Lambda\\
M
\end{pmatrix}$, $m=m_0+t$,
correspond to the Lax pair (\ref{Laxgen2DTL}), where the coefficients 
in the first Lax-Sato equation can be transformed to the form (\ref{Laxgen2DTL})
by taking its expansion at $\lambda=0$, and the system (\ref{gen2DTL}) arises as
a compatibility condition.

Lax-Sato equations (\ref{Hi1},\ref{Hi2}) define the evolution of the series
$\Lambda^\text{in},\Lambda^\text{out}$, $M^\text{in},M^\text{out}$. The only term
containing an interaction between $\Lambda$ and $M$ is $\{\Lambda,M\}$.
The condition $\{\Lambda,M\}=1$ splits out equations for $\Lambda$ and
reduces the hierarchy
(\ref{Hi1},\ref{Hi2}) to the d2DTL hierarchy, while the condition $\Lambda=\ln \l$ --
to the hierarchy, considered by Mart\'{i}nez Alonso and Shabat
\cite{MS02,MS04}, see also Pavlov \cite{Pavlov03}.
\subsection{The dressing scheme}
A dressing scheme for the generalized two-component d2DTL hierarchy can be formulated
in terms of the two-component nonlinear Riemann-Hilbert problem on the unit circle $S$
in the complex plane of the variable $\l$,
\bea
\Lambda^\text{out}=F_1(\Lambda^\text{in},M^\text{in}),
\nn\\
M^\text{out}=F_2(\Lambda^\text{in},M^\text{in}),
\label{Riemann}
\eea
where the functions 
$\Lambda^\text{out}(\l,\mathbf{x},\mathbf{y},t)$, 
$M^\text{out}(\l,\mathbf{x},\mathbf{y},t)$ 
are defined inside the unit circle,
the functions $\Lambda^\text{in}(\l,\mathbf{x},\mathbf{y},t)$, 
$M^\text{in}(\l,\mathbf{x},\mathbf{y},t)$ 
outside the
unit circle by the series of the form (\ref{Lform}), (\ref{Mform}),
with $\Lambda^\text{in}-\ln\lambda$, $M^\text{in}-M_0^\text{in}$ analytic
in the unit disc, and 
$\Lambda^\text{out}-\ln\lambda$, $M^\text{out}-M_0^\text{out}$ analytic
outside the unit disc and decreasing at infinity.
The functions $F_1$, $F_2$ are suggested to define (at least locally) a
diffeomorphism of the plane, $\mathbf{F}\in\text{Diff(2)}$, and we call them
the dressing data. Let us consider a differential form
\beaa
\Omega=\d\Lambda\wedge\d M.
\eeaa
The condition for this form on the unit circle is determined by the Jacobian of
the diffeomorphism defined by $F_1$, $F_2$,
\be
\Omega^\text{out}=
\left|\frac{D(F_1,F_2)}{D(\Lambda^\text{in},M^\text{in})}\right|\Omega^-,
\label{Omega}
\ee
where for the Jacobian we use a notation 
\beaa
\left|\frac{D(f,g)}{D(x,y)}\right|=\det \frac{D(f,g)}{D(x,y)}=
\det
\begin{pmatrix}
\frac{\p f}{\p x}&\frac{\p f}{\p y}\\
\frac{\p g}{\p x}&\frac{\p g}{\p y}
\end{pmatrix}.
\eeaa
Expressing the differential $\d$ in terms of independent variables 
$\l$, $\mathbf{x}$, $\mathbf{y}$, $t$
(\ref{diff}), we come to the conclusion that all the coefficients of the differential
two-form $\Omega$ in terms of these variables transform according to the condition 
(\ref{Omega}).
Normalizing the form by one of the coefficients, we obtain the differential form
continuous on the unit circle. Taking the coefficient corresponding to $d\l\wedge \d t$,
we obtain the relation
\bea 
\left(\left|\frac{D(\Lambda,M)}{D{(\l,t)}}\right|^{-1}\Omega\right)^\text{out}
=\left(\left|\frac{D(\Lambda,M)}{D{(\l,t)}}\right|^{-1}\Omega\right)^\text{in},
\label{formrelation}
\eea
which is equivalent to the generating relation (\ref{genToda}) 
after multiplication by $\lambda^{-1}$,
$$
J_0=\lambda\left|\frac{D(\Lambda,M)}{D{(\l,t)}}\right|=\det
\begin{pmatrix}
\l\p_\l \Lambda & \p_t \Lambda\\
\l\p_\l M       & \p_t M
\end{pmatrix}.
$$
\subsection{Differential reductions}
Recently we have introduced a class of reductions of the Manakov-Santini hierarchy
\cite{LVB2} connected with the interpolating system \cite{Dun08}. Similar reductions 
can be constructed for the generalized two-component d2DTL hierarchy, we are
going to study them in detail elsewhere. Here we will only present the simplest
reduction, which is analogous to the reduction of the Manakov-Santini system 
leading to the interpolating system \cite{Dun08}. The reduced hierarchy is 
defined by the relation
\beaa
(\exp(-\alpha \Lambda)\d\Lambda\wedge\d M)^\text{out}=
(\exp(-\alpha \Lambda)\d\Lambda\wedge\d M)^\text{in},
\eeaa
where $\alpha$ is a parameter ($\alpha=0$ corresponds to the case of d2DTL hierarchy),
which implies
that 
\beaa
J_0=\lambda^{-\alpha}\exp(\alpha \Lambda),
\eeaa
and in terms of the system (\ref{gen2DTL}) we get a reduction
\beaa
\e^{\alpha\phi}=m_t.
\eeaa
This reduction makes it possible to rewrite the system (\ref{gen2DTL}) as
one equation for $m$,
\beaa
m_{tt}=(m_t)^{\frac{1}{\alpha}}(m_{ty}m_x-m_{xy}m_t),
\eeaa
or in the form of deformed d2DTL equation,
\beaa
&&
(\mathrm{e}^{-\phi})_{tt}=m_t\phi_{xy}-m_x\phi_{ty},
\nn\\
&&
m_t=\e^{\alpha\phi}.
\eeaa
\section{Transformations and a symmetric generalization}
The system (\ref{gen2DTL}) we have introduced above doesn't 
preserve the symmetry of the dispersionless 2DTL equation (\ref{d2DTL})
with respect to the variables $x$, $y$. To introduce a symmetric generalization
of the equation (\ref{d2DTL}) and its elliptic version
\bea
(\mathrm{e}^{-\phi})_{tt}=\phi_{z\bar z},
\label{d2DTLell}
\eea
it is possible to change the form of the series for $\Lambda$, $M$ to have an explicit
symmetry between zero and infinity in the complex plane of the spectral variable $\l$,
then the generating relation (\ref{genToda}) will lead to symmetric Lax-Sato equations.
However, we prefer to consider first the transformations of the hierachy that will allow
us to transfer to the symmetric case and will give the connection between different
generalizations of the d2DTL equation.

First, there is a gauge transformation, present already in d2DTL case \cite{TT91,TT95},
which changes the Lax pair,
but preserves the equations
$$
\l\rightarrow\l \exp(-\epsilon\phi),
$$
where $\epsilon$ is a parameter. After this transformation we get $\Lambda$
of the form
\beaa
&&
\Lambda^\text{out}=\ln\lambda-\epsilon\phi + \sum_{k=1}^{\infty}l^+_k\lambda^{-k},\\
&&
\Lambda^\text{in}=\ln\lambda+(1-\epsilon)\phi+\sum_{k=1}^{\infty}l^-_k\lambda^{k}.
\eeaa
In the Lax pair one should perform a substitution
\beaa
&&
\l\rightarrow\l \exp(-\epsilon\phi),
\;\partial_\l\rightarrow \exp(\epsilon\phi)\partial_\l.
\\&&
\partial_x\rightarrow \partial_x + \epsilon\lambda \phi_x \partial_\lambda,
\;
\partial_y\rightarrow \partial_y + \epsilon\lambda \phi_y \partial_\lambda,
\;
\partial_t\rightarrow \partial_t + \epsilon\lambda \phi_t \partial_\lambda,
\eeaa
In the elliptic d2DTL case (\ref{d2DTLell}) 
for $\epsilon=\frac{1}{2}$ we get a symmetric Lax pair
\beaa
&&
\p_z\mathbf\Psi=L_1\mathbf\Psi=\left((\lambda \e^{-\frac{1}{2}\phi})\partial_t + 
{\frac{1}{2}}(\phi_z + \lambda\e^{-\frac{1}{2}\phi}\phi_t)
\l\partial_\lambda\right)\mathbf\Psi,\\
&&
\p_{\bar z}\mathbf\Psi=L_2\mathbf\Psi=
\left((\frac{1}{\l}\e^{-\frac{1}{2}\phi})\partial_t - 
{\frac{1}{2}}(\phi_{\bar{z}} + \lambda\e^{-\frac{1}{2}\phi}\phi_t)
\l\partial_\lambda\right)\mathbf\Psi,
\eeaa
and on the unit circle $L_1=\bar L_2$.

To get a symmetric two-component generalization of the ellyptic d2DTL equation
and a symmetric Lax pair for it, we should also use a hodograph type transformation
$$t=\tau- \alpha m_0$$
(where $\tau$ is a new `time', $\alpha$ is a parameter), which
gives $M$ of the form 
\beaa
&&
M^\text{out}=
M_0^\text{out} +(1-\alpha)m_0+\sum_{k=1}^{\infty}m^+_k\e^{-k\Lambda^{+}},\\
&&
M^\text{in}=M_0^\text{in} - \alpha m_0+\sum_{k=1}^{\infty}m^-_k\e^{k\Lambda^{+}},\\
&&
M_0=\tau+x\e^\Lambda+y\e^{-\Lambda}+\dots
\eeaa
Derivatives transform as follows,
\beaa
\partial_x\rightarrow \partial_x+\frac{\alpha{m_0}_x}{1-\alpha{m_0}_\tau}\partial_\tau,\;
\partial_y\rightarrow \partial_y+\frac{\alpha{m_0}_y}{1-\alpha{m_0}_\tau}\partial_\tau,\;
\partial_t\rightarrow \partial_\tau+
\frac{\alpha{m_0}_\tau}{1-\alpha{m_0}_\tau}\partial_\tau.
\eeaa 
Applying these transformations to the system (\ref{gen2DTL}), where $m=m_0+t$,
we obtain a one-parametric
family of two-component generalizations of the d2DTL equation.

Taking $x=z$, $y=\bar z$, $\epsilon=\frac{1}{2}$, $\phi\rightarrow -2\varphi$,
$\alpha=\frac{1}{2}$, $m_0=-2\mathrm{i} \mu$,
we get
\beaa
&&
\Lambda^\text{out}=\ln\lambda+\varphi+\sum_{k=1}^{\infty}l^+_k\lambda^{-k},\quad
\Lambda^\text{in}=\ln\lambda-\varphi+\sum_{k=1}^{\infty}l^-_k\lambda^{k},\\
&&
M^\text{out}=M_0^\text{out} +
\mathrm{i} \mu+\sum_{k=1}^{\infty}m^+_k\e^{-k\Lambda},\quad
M^\text{in}=M_0^\text{in} -\mathrm{i} \mu+\sum_{k=1}^{\infty}m^-_k\e^{k\lambda},\\
&&
M_0=\tau + z\e^\Lambda+\bar z\e^{-\Lambda}+\dots
\eeaa
where for the case of ellyptic d2DTL we suggest that $\mu$, $\varphi$ are real,
and on the unit circle $\lambda\bar\lambda=1$
\beaa
&&
M^\text{out}=\bar M^\text{in},\\
&&
\Lambda^\text{out}=-\bar\Lambda^\text{in}.
\eeaa
From the Lax pair (\ref{Laxgen2DTL}) we obtain a symmetric Lax pair
\beaa
&&
\partial_{z}\mathbf\Psi=L_1\mathbf\Psi,\quad L_1=(\lambda \e^\varphi u + v)\partial_\tau + 
((\varphi_\tau v - \varphi_z)- \lambda u \e^\varphi \varphi_\tau)
\l\partial_\lambda,\\
&&
\partial_{\bar z}\mathbf\Psi=L_2\mathbf\Psi,
\quad L_2=(\frac{1}{\lambda} \e^\varphi \bar u + \bar v)\partial_\tau - 
((\varphi_\tau \bar v - \varphi_{\bar z})- \frac{1}{\l}\bar u \e^\varphi \varphi_\tau)
\l\partial_\lambda,
\eeaa
on the unit circle $L_1=\bar L_2$,
\beaa
&&
u=\frac{1}{1+\mathrm{i} \mu_\tau},
\quad v=\frac{-\mathrm{i} \mu_z}{1-\mathrm{i}\mu_\tau}.
\eeaa
Equation (\ref{gen2DTL}) transforms to the symmetric two-component generalization
of the ellyptic d2DTL equation (\ref{d2DTLell}),
\beaa
&&
(v_{\bar z}+ \e^\varphi u \partial_\tau (\e^\varphi \bar u) + v \partial_\tau \bar v)
-\text{c.c.}=0,\\
&&
(\partial_{\bar z}(\varphi_\tau v - \varphi_z) + 
\e^\varphi u \partial_\tau(\bar u \e^\varphi\varphi_\tau)
-v\partial_\tau(\varphi_\tau\bar v - \varphi_{\bar z}) +
u \bar u \e^{2\varphi}\varphi_\tau\varphi_\tau)
\\&&\qquad\qquad
+\text{c.c.}=0
\eeaa
If $\mu=0$ ($u=1$, $v=0$), the first equation vanishes, the second gives 
the d2DTL equation for
$\phi=(-2\varphi)$.

If $\varphi=0$,  the second equation vanishes, the first gives
$$
(v_{\bar z}+  u \partial_\tau (\bar u) + v \partial_\tau \bar v)
-\text{c.c.}=0,
$$
or, in explicit form,
$$
\mu_{\tau\tau}=\frac{1}{2}(\mu_z\mu_{\bar{z}}-(1+\mu_\tau^2))^{-1}
(\mu_\tau^2(\partial_\tau(\mu_z\mu_{\bar{z}})-
\text{i}(\mu_{z\tau}\mu_{\bar{z}}-\mu_z\mu_{\bar{z}\tau}))
-\mu_{z{\bar{z}}}(1+\mu_\tau^2)).
$$
\section*{Acknowledgments}
The author is grateful to S.V. Manakov and P.M. Santini for
useful discussions.
This research was partially supported by the Russian Foundation for
Basic Research under grants no. 10-01-00787, 09-01-92439, 
and by the President of Russia
grant 4887.2008.2 (scientific schools).

\end{document}